\documentclass[12pt]{article}

\usepackage{amsmath}
\usepackage{amsfonts}
\usepackage{pgf}
\usepackage[numbers]{natbib}

\renewcommand{\theequation}{\thesection.\arabic{equation}}

\title{Machine Learning Assisted Adjustment Boosts Efficiency of Exact Inference in Randomized Controlled Trials}

\author{Han Yu, Alan D. Hutson, Xiaoyi Ma
\thanks{Department of Biostatistics and Bioinformatics, Roswell Park Cancer Institute, Elm and Carlton Streets, Buffalo, NY 14623. This work was supported by Roswell Park Cancer Institute and National Cancer Institute (NCI) grant P30CA016056,  NRG Oncology Statistical and Data Management Center grant U10CA180822 and IOTN Moonshot grant  U24CA232979-01, ARTNet Moonshot grant U24CA274159-01, CAP-IT grant U24CA274159-02. }} 
\date{}

\begin{document}
\maketitle

\begin{abstract}

In this work, we proposed a novel inferential procedure assisted by machine learning based adjustment for randomized control trials. The method was developed under the Rosenbaum's framework of exact tests in randomized experiments with covariate adjustments. Through extensive simulation experiments, we showed the proposed method can robustly control the type I error and can boost the statistical efficiency for a randomized controlled trial (RCT). This advantage was further demonstrated in a real-world example. The simplicity, flexibility, and robustness of the proposed method makes it a competitive candidate as a routine inference procedure for RCTs, especially when nonlinear association or interaction among covariates is expected. Its application may remarkably reduce the required sample size and cost of RCTs, such as phase III clinical trials.

\end{abstract}

\renewcommand{\theequation}{\thesection.\arabic{equation}}
\pagebreak

\section{Introduction} \label{sec:intro}

The randomized controlled trial (RCT) is the gold standard in determining treatment efficacy. RCTs are characterized by their extensive participant involvement, which typically encompass a substantial cohort size, particularly evident in phase III clinical trials. However, this rigorous process comes at a considerable cost. Especially for trials on immunotherapy, the cost of some of the newest therapies can reach \$850,000 per patient when including the value of medical support necessary to deliver these treatments \cite{schaft2023future}.  Given the significant investment required, maximizing the statistical power of RCTs becomes critical. 

A common approach to improve the statistical efficiency is baseline adjustment using linear models, such as analysis of covariance (ANCOVA). Related models and their applications in RCT has been extensively discussed \cite{ye2023toward, van2024covariate}. The method of covariate adjustments is supported by FDA in phase III oncology trials due to the enhanced efficiency and its minimal impact on bias or type I error rate \cite{fda2021}. Hence, the agency advises sponsors to incorporate adjustments for covariates expected to show the strong association with the outcome of interest. In 2002, Rosenbaum proposed method of exact inference which is free of distributional assumptions \cite{rosenbaum2002covariance}. This method is endorsed by the FDA guidance, which states that "Sponsors can conduct randomization/permutation tests with covariate adjustment (Rosenbaum 2002)". Although this method was based on ordinary least square (OLS) linear regression, the author raised possibility of using more general forms of covariate adjustment.  

Nowadays, due to the ever-decreasing cost of medical testing and sequencing techniques, the number of baseline data available in clinical trials has been growing dramatically. The rich data can provide a much finer picture of individuals and thus capture larger variation in the outcomes. This makes it possible to design inferential procedures with superior efficiency. However, the traditional adjustment methods using OLS method quickly break down when the number of covariates increases to a relatively large scale. Even if penalized regressions \cite{hoerl1970ridge, tibshirani1996regression, chen2016xgboost, zou2005regularization} can accommodate high-dimensional data, such linear models are neither capable of handling nonlinear associations nor complex interactions.

On the other hand, machine learning techniques have been successfully applied in numerous settings during the last decade. Supervised learning methods such as random forests (RFs), boosting machines, neural networks \cite{breiman2001random, chen2016xgboost, hastie2009elements, lecun2015deep} are known to be capable of handling high-dimensional data and model complex functional forms. While efforts have been made towards the application of machine learning models for the covariates adjustment in randomized studies, most of the works focused on estimations and asymptotics \cite{wager2016high, opper2021improving, williams2022optimising}. In this work, we propose a novel approach of machine learning based covariate adjustment under the Rosenbaum's framework, which makes exact inference possible \cite{rosenbaum2002covariance}. The proposed method focuses on continuous outcomes, and is shown to be an unbiased and flexible adjustment method to boost the statistical efficiency for a RCT. These characteristics make it a novel and powerful inferential procedure for RCTs, especially when nonlinear association or between-covariate interaction is expected. The RF model is used under this framework as a proof of concept, but it can be extended to more general modeling approaches. Its application may remarkably reduce the required sample size and cost of RCTs, especially for phase III clinical trials.

\section{Methods} \label{sec:methods}
\subsection{Hypothesis testing under non-parametric adjustment}
In this work, we focus on RCTs with continuous outcomes and with an objective to compare two group means. Our method is based on the Rosenbaum's framework. For details, please see \citet{rosenbaum2002covariance}. As a brief background, suppose there are $n$ subjects, $j = 1, \dots, n$, and the response of subject $j$ would be $r_{T_j}$ or $r_{C_j}$ if $j$ were assigned to treatment or control, respectively \cite{splawa1990application, rubin1974estimating}. The treatment effect can be written as $\tau_i = r_{T_j} - r_{C_j}$, and it can never be observed because subject $j$ can only receive one of the treatments. The treatment effect is \textit{additive} if the treatment causes the response to change by a fixed amount $\tau$ for every $j$. The covariate vector $x_j$ describes the baseline characteristic of $j$. The variable $Z_j$ is the treatment indicator, where $Z_j=1, 0$ indicates $j$ is assigned to the treatment  or control group, respectively. Thus, the response of $j$ can be written as $R_j=Z_jr_{T_j} + (1-Z_j)r_{C_j}$. Following Rosenbaum \cite{rosenbaum2002covariance}, we denote $\mathbf{Z}=(Z_1, Z_2, \dots, Z_n)^T$, $\mathbf{R}=(R_1, R_2, \dots, R_n)^T$, $\mathbf{r}_c=(r_{C_1}, r_{C_2}, \dots, r_{C_n})^T$ and use $\mathbf{X}$ as the covariate matrix whose $j$th row is $\mathbf{x}_j^T$. The fixed but unobserved vector $\mathbf{r}_C$ are potential outcomes of all subjects if assigned to the controlled group, which can be expressed as $\mathbf{r}_C=\mathbf{R} - \tau\mathbf{Z}$. Under this framework, the only stochastic component is the random assignment $\mathbf{Z}$. The test of $H_0:\tau=\tau_0$ can be performed using the statistic $t(\mathbf{Z}, \mathbf{Y})$, where $\mathbf{Y} = \mathbf{R} - \tau_0\mathbf{Z}$.  The response variable $\mathbf{Y}$ is the observed outcome $\mathbf{R}$ adjusted with respect to $\tau_0$, which will be the variable used in the inference procedure. Thus we have $\mathbf{Y} = \mathbf{R}$ when $\tau_0=0$. Under this framework, the exact inference is the randomization inference derived from the randomization distribution of the statistics. The parametric distributions such as the normal distribution are used as approximations to randomization distributions \cite{rosenbaum2002covariance}.

When covariate information $\mathbf{X}$ is available and associated with the response variable $\mathbf{Y}$, then we can write $\mathbf{Y}=g(\mathbf{X})+\mathbf{e}$. The residual $\mathbf{e}$ is expected to have less variation than $\mathbf{Y}$ if the baseline covariates explain part of the variation, %Note that if $H_0$ is true, we have $\mathbf{r}_C = \mathbf{R} - \tau_0\mathbf{Z}$,
and it can be used in place of $\mathbf{Y}$ for inference. Thus the test statistic can be modified to $t(\mathbf{Z}, \mathbf{e})$, which can lead to an improvement in efficiency due to smaller variation in $\mathbf{e}$.

Linear models are natural choice of $g$, but they tend to fail when the association is complex or the data is high dimensional. Here we investigate whether using modern machine learning method will achieve better efficiency for testing $H_0:\tau=\tau_0$ as compared to the straight two group comparison, while controlling the type I error rate properly. In this work, we select RF as the exact inference approach. 

The RF approach was selected for our approach due to its minimal requirement of parameter tuning and invariance to monotonic transformation of features \cite{breiman2001random}. More importantly, the out-of-bag (OOB) predictions can be used to attenuate the overfitting, thus avoiding manually splitting the data into training and validation set. Such simplicity makes it ideal for RCTs, which typically requires detailed preplanning. 

The proposed procedure for testing $H_0:\tau=\tau_0$ is straight forward:
\begin{enumerate}
	\item Use the data from all subjects and build a RF model using $\mathbf{X}$ as the features and $\mathbf{R - \tau_0\mathbf{Z}}$ as the response $\mathbf{Y}$. 
	\item Obtain the OOB predictions for all subjects and calculate the residuals $\mathbf{e}$.
	\item Perform the permutation test, Wilcoxon rank-sum test or two sample $t$-test based on $\mathbf{e}$ and $\mathbf{Z}$.
	\item Estimate the treatment effect, standard error or confidence interval by treating $\mathbf{e}$ as the outcome variable.
\end{enumerate}

\subsection{Estimation}

Without loss of generality, we focus on the null hypothesis $H_0: \tau_0=0$ and relax the additivity assumption for the discussion of estimation. When $H_0: \tau_0=0$, we have $\mathbf{Y} = \mathbf{R}$ which is the observed response. Conceptually, the adjustment by $g(\mathbf{X})$ shifts the response at any point in the covariate space by the same amount for both arms. Therefore, the difference in the residuals should be the same as that in the original response, so the treatment effect is estimated as, 
\begin{equation}
	\label{eqn:est}
	\begin{aligned}
		\hat{\tau} & = \frac{1}{n_1}\sum_{Z_i=1}e_i-\frac{1}{n_0}\sum_{Z_i=0}e_i  
		%& = \frac{1}{n_1}\sum_{Z_i=1}(Y_i-g(X_i))-\frac{1}{n_0}\sum_{Z_i=0}(Y_i-g(X_i)) \\
		= \frac{1}{n_1}\sum_{i=1}^n(Y_i-g(X_i))Z_i-\frac{1}{n_0}\sum_{i=i}^n(Y_i-g(X_i))(1-Z_i), \\
	\end{aligned}
\end{equation}
where $n_0$ and $n_1$ are numbers of subjects in control and treatment groups. The same estimator has been proposed by Opper \cite{opper2021improving} for a different inference procedure. The condition for $\hat{\tau}$ in Equation \ref{eqn:est} to be unbiased estimator was also given \cite{opper2021improving}. In our framework, the estimation of $g(\mathbf{X})$ does not involve treatment assignment, so it is invariant under randomization. Therefore, the only stochastic component is $Z_i$. Therefore, we can write $Y_i = r_{C_i} + Z_i\tau_i, i=1,2,\dots n$, then the expectation of estimated treatment effect is given as,
 \begin{equation}
 	\begin{aligned}
 		E(\hat{\tau}) & =  E( \frac{1}{n_1} \sum_{i=1}^n(Y_i-g(X_i))Z_i-\frac{1}{n_0}\sum_{i=i}^n(Y_i-g(X_i))(1-Z_i) )\\
 		& =  \frac{1}{n}\sum_{i=1}^n\tau_i - \frac{1}{n}\sum_{i=1}^n(E(g(X_i)|_{Z_i=1}) - E(g(X_i)|_{Z_i=0})).\\
 	\end{aligned}
 \end{equation}
Under additivity, we have $\tau_i=\tau, i=1, 2, \dots, n$. Thus, $E(\hat{\tau}) =  \tau + \frac{1}{n}\sum_{i=1}^n(g(X_i)|_{Z_i=1} - g(X_i)|_{Z_i=0})$. Therefore, a sufficient condition for $\hat{\tau}$ to be an unbiased estimator, i.e., $E(\hat{\tau}) =  \tau$ is 
$$E(g(X_i)|_{Z_i=0}) = E(g(X_i)|_{Z_i=1}).$$ 
One would assume that this condition is natually fulfilled since the estimation of $g$ does not involve treatment assignment. However, it can be violated particularly if $g$ overfits the data. Overfitting occurs when $g(X_i)$ is overly close to the observed response $Y_i$, meaning that as the flexibility of $g$ increases, we have $g(X_i)\to r_{C_i}$ when $Z_i=0$, and $g(X_i)\to r_{C_i} + \tau_i$ when $Z_i=0$. Consequently, we have $E(g(X_i)|_{Z_i=1}) - E(g(X_i)|_{Z_i=0})\to \tau_i$ and  $E(\hat{\tau}) \to 0$. This suggests when $g$ overfits the data, the estimated treatment effect will be biased towards $0$.
As long as the machine learning model is adequately flexible, it can always overfit the data even if the sample size is large (e.g., the $k$ Nearest Neighbor with $k=1$). Therefore, for models with high complexity, the bias term is $O(1)$ as $n\to\infty$, thus simply increasing the sample size does not necessarily mitigate the bias. One way to counter this problem is to use the predictions from models not trained on the observations at question, which can be easily achieved through cross-validations, or OOB predictions. 

Next, we derive the variance of the permutation distribution of $\hat{\tau}$. Note that $\hat{\tau}$ can be rewritten as,
\begin{equation}
	\begin{aligned}
		\hat{\tau} & = \sum^n_{i=1}\frac{ne_iZ_i}{n_1n_0} - \frac{1}{n} \sum^n_{i=1}e_i.\\
	\end{aligned}
\end{equation}
Recall that in our method, the residuals are fixed after the estimation of $g$. Therefore, the variance of the estimator is given as,
\begin{equation}
	\begin{aligned}
		Var(\hat{\tau}) & = \frac{1}{n_1n_0}\sum^n_{i=1}e_i^2.\\
	\end{aligned}
\end{equation}
The equation above suggests the variance will be smallest when $\sum^n_{i=1}e_i^2$ is minimized, which corresponds to a model $g$ that is optimized towards minimal mean squared error. This observation bridges the inference problem with classic predictive modeling problems. It is notable that Opper obtained the same form of variance in his work\cite{opper2021improving}. However, the author ignored the variability of $g(\mathbf{X})$, and assumed the residuals to be fixed, which does not align with their procedure that requires estimation of $g(\mathbf{X})$ under each randomization. Interestingly, the same expression is correct under our framework, where the residuals are indeed fixed because estimation of $g(\mathbf{X})$ does not depend on $\mathbf{Z}$.

When the treatment effect is non-additive, the treatment effect is defined as $\tau =  \frac{1}{n}\sum_{i=1}^n\tau_i$. Therefore, the above results on unbiasedness still hold. On the other hand, non-additivity implies that there are interactions between treatment and covariates. Thus the function $g(\mathbf{X})$ can be better described as two separate functions for two arms, namely $g_0(\mathbf{X})$ and $g_1(\mathbf{X})$. Although testing the null hypothesis under the non-additivity assumption is not very well defined, a natural question under non-additivity is whether modeling $g_0(\mathbf{X})$ and $g_1(\mathbf{X})$ individually will gain efficiency over estimating a common regression function $g(\mathbf{X})$ by pooling the observations from two arms together. Such a strategy was  adopted by Wager\cite{wager2016high} and Opper\cite{opper2021improving}. To investigate this problem, we examined the estimator of treatment effect from the cross-estimation (CE) method by Wager  \cite{wager2016high},
 \begin{equation}
	\begin{aligned}
\hat{\tau}_{CE} = \frac{1}{n}\sum_{i=1}^n(g_1(X_i)-g_0(X_i)) + \sum_{Z_i=1}\frac{Y_i-g_1(X_i)}{n_1} - \sum_{Z_i=0}\frac{Y_i-g_0(X_i)}{n_0}.
	\end{aligned}
\end{equation}
Here, we ignored the $(-i)$ notation, which was used to denote that the estimator of $g$ was not dependent on the $i$th training ovservation. %For simplication, we assume the knowledge of true $g_1$ and $g_0$ is available so their variances are zeros. 
Then based on their result, the variance of the estimator is given as,
 \begin{equation}
	\begin{aligned}
		%Var(\hat{\tau}) & =  Var( \sum_{Z_i=1}\frac{Y_i-g_1(X_i)}{n_1} - \sum_{Z_i=0}\frac{Y_i-g_0(X_i)}{n_0} )\\
		Var(\hat{\tau}_{CE}) & = \sum_{z\in (0,1)} \sum_{Z_i=z}\frac{(Y_i-(1-p)g_1(X_i)-pg_0(X_i))^2}{n_z(n_z-1)},
	\end{aligned}
\end{equation}
where $n_z= {n \choose n_0}$ is the number of all unique randomizations and $p=n_1/n$. Note that the variance depends on $g_0$ and $g_1$ only through a weighted sum $g(X_i) = (1-p)g_0(X_i)+pg_1(X_i)$, which can be estimated through pooling the observations from two arms and proper weighting. Therefore, even under non-additivity, the knowledge of $g_0$ and $g_1$ does not lead to a further gain in efficiency, so the estimation of $g(\mathbf{X})$ across two components is unnecessary. In fact, under a balanced design, the sample size for training the model will be halved to train the models for the control and treatment groups separately, which can result in reduced model performance and consequently lower inference efficiency. Therefore, the proposed method is not only more straightforward for exact inference, but it also achieves a better efficiency through pooling observations from both intervention arms.

\section{Simulations} \label{sec:simu}

To demonstrate the performance of the proposed method, we conducted a set of comprehensive simulation studies under various scenarios for testing a one-sided hypothesis $H_0: \tau=0$ versus $H_1: \tau>0$. The results can be readily extended to two-sided tests. In each scenario, we generated samples of size $N=50, 100, 200$ and $400$ independent subjects, who were then randomized into two balanced groups. There were $p=40$ mutually independent covariates following standard normal distributions. In each setting 10,000 Monte Carlo simulations were performed. The outcome $y$ was generated using the four different models:
\begin{enumerate}
	\item Model 1 (primarily nonlinear no interaction): $$ y = \tau Z + \beta \sigma(x_1/2) + \beta x_2^2 + \beta cos(x_3) + \beta x_4 + \epsilon, $$ 
	\item Model 2 (primarily nonlinear with interaction): $$ y = \tau Z + \beta \sigma(x_1/2) + \beta x_2^2 + \beta cos(x_3) + \beta sign[cos(x_3)]x_4 + \epsilon, $$
	\item Model 3 (primarily linear no interaction): $$ y = \tau Z + \beta \sigma(x_1/2) + \beta x_2 + \beta x_3 + \beta x_4 + \epsilon, $$
	\item Model 4 (primarily nonlinear with by treatment interaction): $$ y = \tau Z + \beta [\sigma(x_1/2)-1/2]Z + \beta x^2_2 + \beta cos(x_3) + \beta x_4 + \epsilon, $$
\end{enumerate}
where $Z$ is the treatment indicator, $\tau$ is the treatment effect, $\sigma(x)=\exp(x)(1+\exp(x))^{-1}$ is the sigmoid function, $x_k$ is the $k$th covariate, and $\epsilon$ is the error term. The subscript for subject is omitted. We examined $\beta=0.2, 0.5, 0.8$ to represent different levels of association between the covariates and the outcome measure. In addition, we utilized error terms from standard normal, log-normal, and Gumbel distributions. For comparing the outcomes between the two treatment ars we utilized the Wilcoxon rank-sum test and the two-sample $t$-test. We also included $t$-test and Wilcoxon rank-sum test on the residuals from linear regression on all baseline covariates ($t$-test-LM and Wilcoxon-LM). Finally, we included the proposed RF adjusted methods, which use the $t$-test and Wilcoxon rank-rum test on the residuals from RF model ($t$-test-RF and Wilcoxon-RF). The Wager's cross-estimation (CE) method was also included for comparison, where a one-sided $Z$-test was constructed based on the estimate and standard error from the \texttt{ate.randomForest} function in R \texttt{crossEstimation} package \cite{wager2016high}.  We examined the type I error control with $\tau=0$, and the statistical power with $\tau=0.3$ and $0.6$. The results of all simulations are shown in the supplementary materials.

\begin{figure}[!htb]
	\centering
	\includegraphics[width=4in]{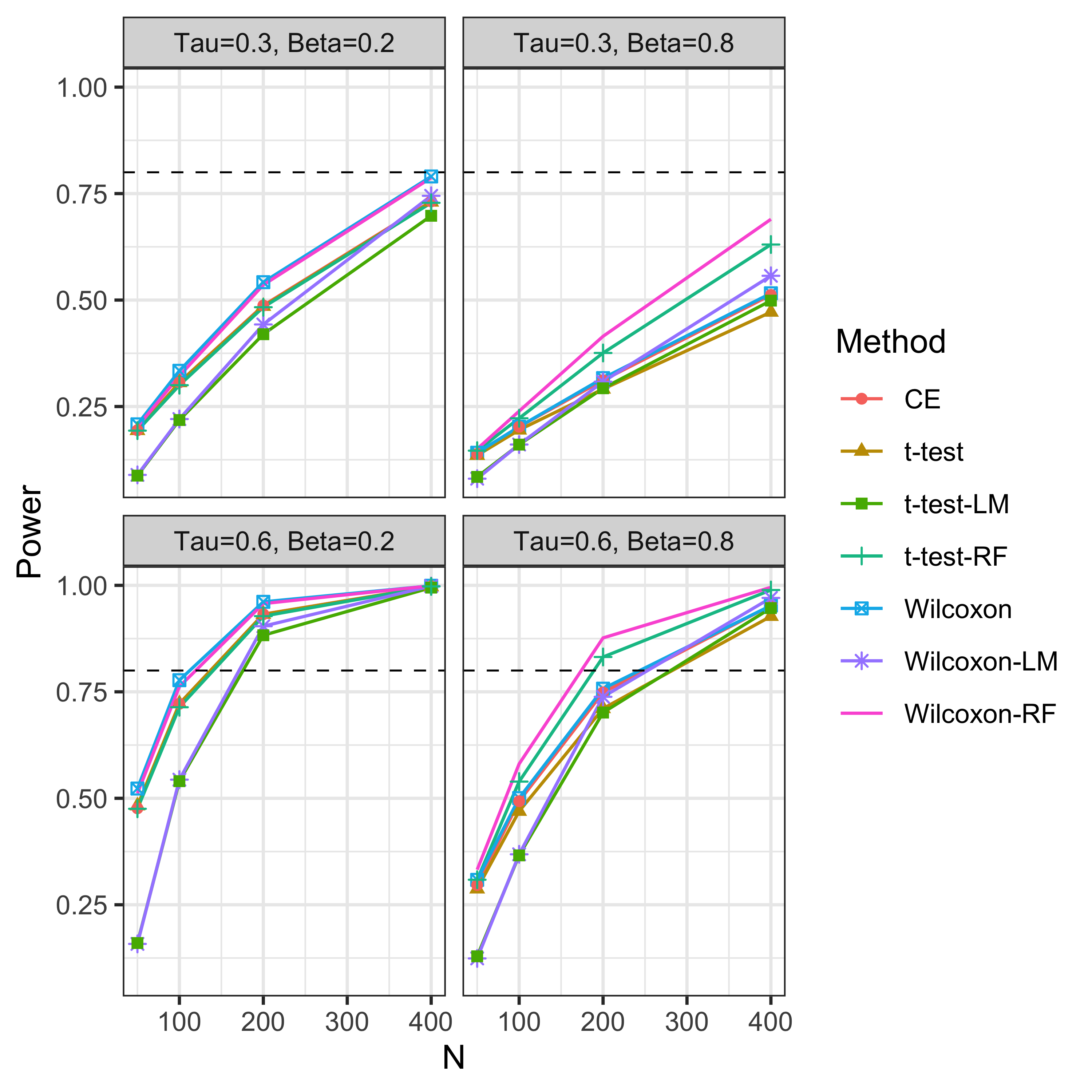}
	\caption{The statistical power of the testing procedures under different settings of Model 1 and with error terms from Gumbel distributions. Dashed line indicates power of 0.8.}
	\label{fig:simu_2}
\end{figure}

\begin{figure}[!htb]
	\centering
	\includegraphics[width=4in]{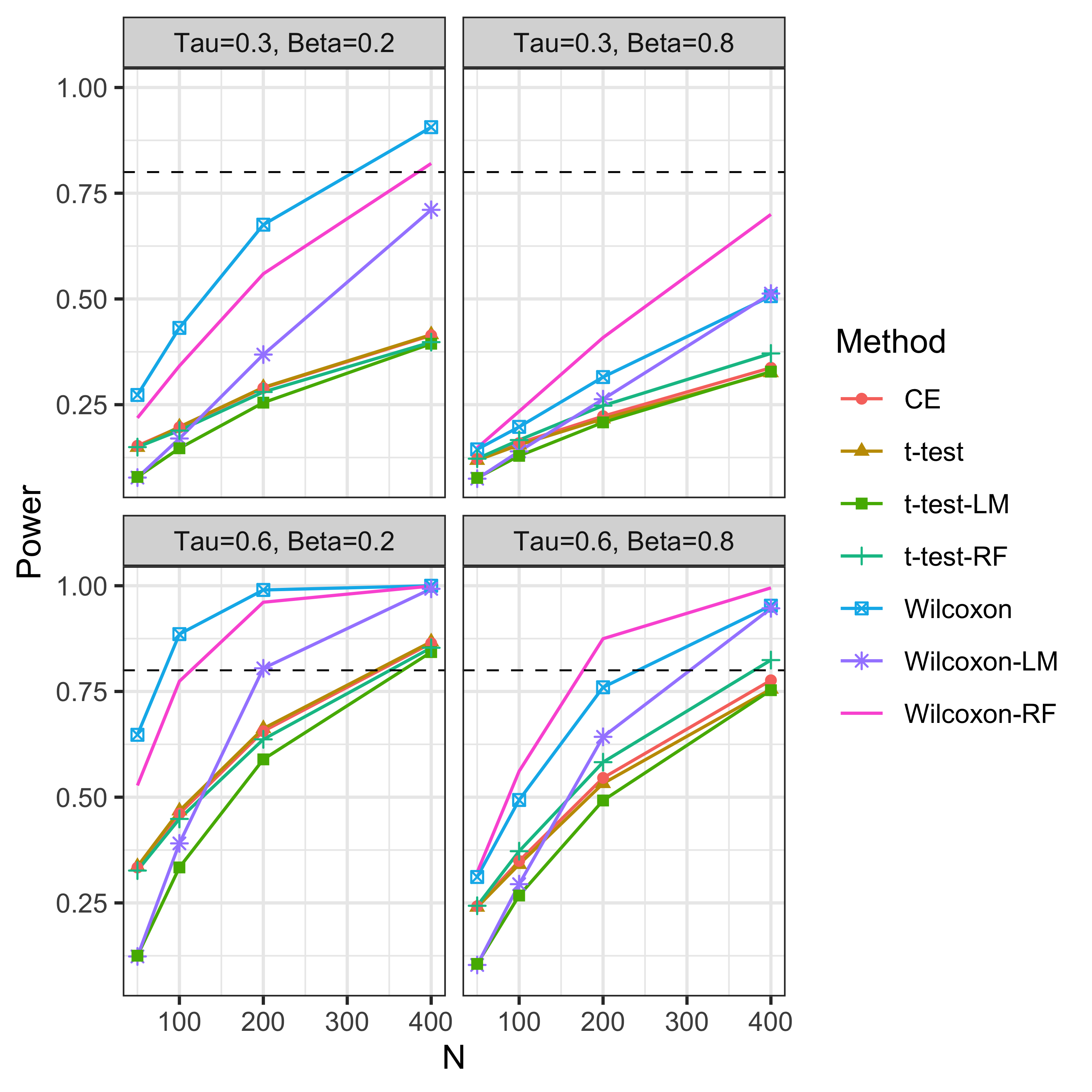}
	\caption{The statistical power of the testing procedures under different settings of Model 1 and with error terms from log-normal distributions. Dashed line indicates power of 0.8.}
	\label{fig:simu_3}
\end{figure}

Figures \ref{fig:simu_1} shows that all tests robustly control the type I error rates when $\tau=0$ under non-Gaussian error terms. In models with primary nonlinear associations, the RF adjusted tests ($t$-test-RF and Wilcoxon-RF) show higher power than the other methods. In the right panel of Figure \ref{fig:simu_2}, we see that the RF tests attains over 80\% power at around $N=170$, while the other tests require approximately 80 more subjects to achieve the same power. In models with primary linear associations, the performance of RF tests is mostly comparable with that of the linear adjusted methods. Only when the  association is very strong ($\beta=0.8$), the linear model methods will have a advantage of around 7\% (Figures S3 \& S7). It should be noted that under some scenarios (e.g. small sample sizes), the linear model approach tends to hurt the efficiency, having power lower than the straight $t$-test and Wilcoxon test, even when the true association is linear. On the other hand, the RF tests have similar or better performance than the unadjusted tests in most cases. One exception is the setting with weak associations and log-normal errors (Figure \ref{fig:simu_3}). However, as the association becomes stronger, the RF test still outperforms other methods. In addition, although the Rosenbaum's framework assumes an additive treatment effect, the proposed method shows equally well type I error control and efficiency improvement when there are interactions between treatment and covariates (Figures S4, S8, S12). It is also notable that our proposed method consistently outperforms the CE method \cite{wager2016high}, even when the covariate by treatment interaction is present. This demonstrates that the explicit modeling of $g$ separately for two arms using nonparametric methods reduces statistical efficiency due to the smaller training sample sizes. The results are similar when the randomization is unbalanced (data not shown). Overall, our simulations suggest that, in most cases, the Wilcoxon-RF test as the best option among the methods tested.

\begin{figure}[!htb]
	\centering
	\includegraphics[width=4in]{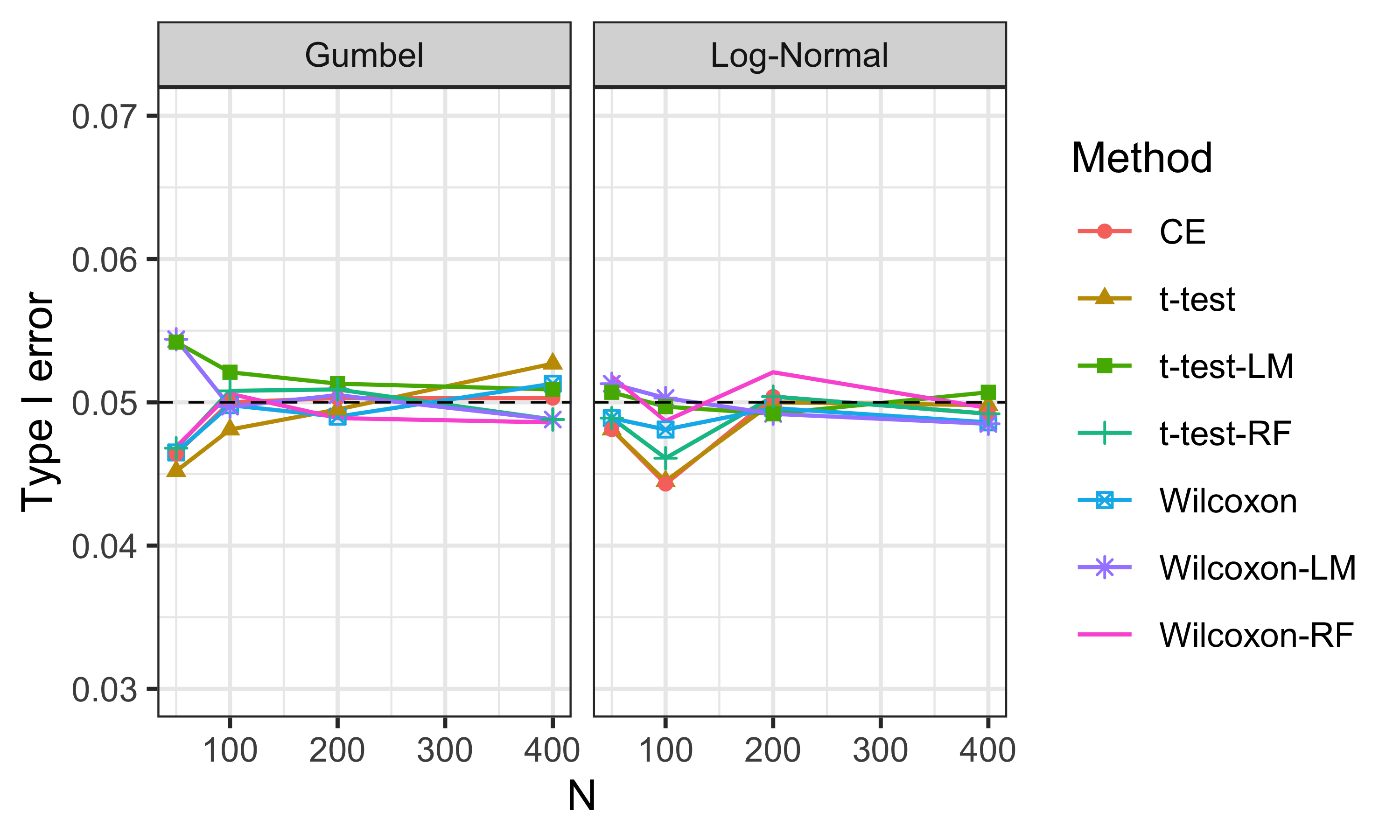}
	\caption{Type I error of the testing procedures under Model 1 with $\tau=0, \beta=0.8$ and error terms from Gumbel and log-normal distributions. Dashed line indicates level of 0.05.}
	\label{fig:simu_1}
\end{figure}

\section{Diet Intervention Data Example} \label{sec:examples}

\begin{figure}[!htb]
	\centering
	\includegraphics[width=2.5in]{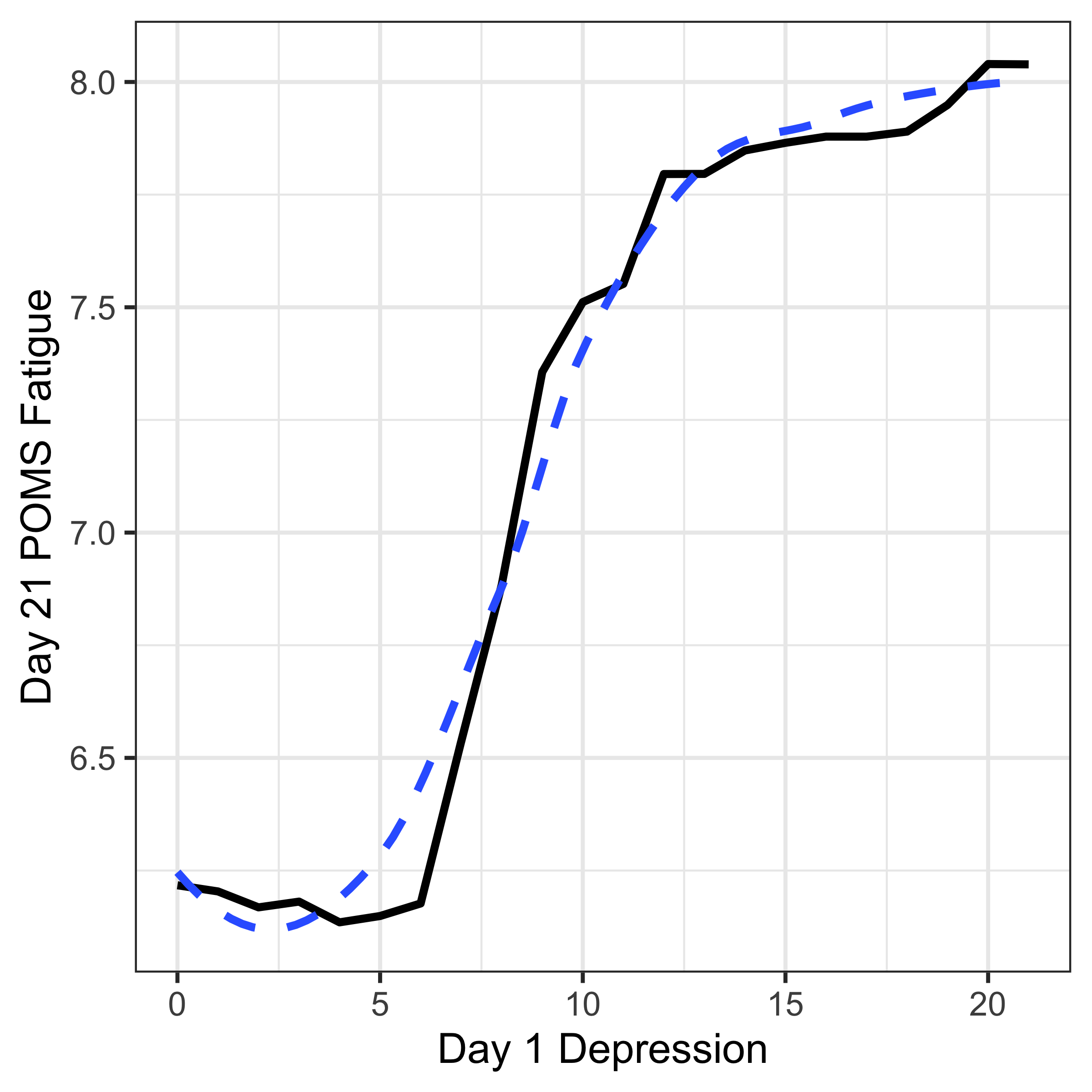}
	\caption{The partial dependence of Day 21 fatigue score and baseline depression score. Blue dashed line is the smoothed partial dependence by Loess smoother.}
	\label{fig:example}
\end{figure}

In this example, the investigators studied whether a brief diet intervention can reduce the symptoms of depression in young adults. There were 76 participants randomly allocated to a brief 3-week diet intervention (Diet Group) or a habitual diet control group (Control Group), with $n=38$ per group. The primary and secondary outcome measures were assessed at both baseline and after the intervention, which included symptoms of depression (Centre for Epidemiological Studies Depression Scale, CESD-R; and Depression Anxiety and Stress Scale-21 depression subscale, DASS-21-D), current mood (Profile of Mood States, POMS), self-efficacy (New General Self-Efficacy Scale, GSES) and memory (Hopkins Verbal Learning Test). There are a total of 23 baseline variables that can be used as covariates. The GSES score were log-transformed as specified in the article. The ANCOVA was used for the analysis of both primary and secondary outcomes with the baseline scores adjusted as covariates. The results showed that the Diet group had significantly lower self-reported depression symptoms than the Control Group on the CESD-R and DASS-21 depression subscale. However, none of the POMS scores showed significant difference. Here we used the Wilcoxon-RF method to re-analyze the difference in POMS fatigue scores between two groups. Two-sided tests were used at $\alpha=0.05$. The results in Table \ref{tab:example} show that only the Wilcoxon-RF shows a significant difference in the post-intervention fatigue scores between two groups ($p=0.0451$). This demonstrates the higher statistical power of the proposed method. Based on the residuals, the final estimated treatment effect is -1.57 (Diet vs. Control), with 95\% confidence interval of (-3.01, -0.05). The variable importance from RF model suggests baseline depression is the most important predictor. The partial dependence plot (Figure \ref{fig:example}) suggests the post-intervention fatigue does not change with the baseline depression within the range of $0 \sim 5$, but starts to show a linear trend in the range of $6\sim 12$, and then plateaus. The sigmoid shape of what may not be well described by a linear model, but can be better captured by a nonlinear model like RF, which explains why our method achieves better efficiency.

\begin{table}
	\centering
	\begin{tabular}{c | c} 
		\hline
		Method & $p$-value \\ [0.5ex] 
		\hline
		CE & 0.2505\\
		$t$-test    & 0.2907 \\ 
		Wilcoxon    & 0.2452 \\
		$t$-test-LM & 0.3112 \\		
		Wilcoxon-LM & 0.3661 \\		
		$t$-test-RF & 0.1248 \\		
		Wilcoxon-RF & 0.0451 \\ [1ex] 
		\hline	
	\end{tabular}
	\caption{Results of testing methods applied to the diet intervention data for comparing post-intervention POMS fatigue score between the Diet and Control groups. All tests are two-sided.}		
	\label{tab:example}
\end{table}

\section{Discussion} \label{sec:discussion}

In this work, we presented a novel inferential procedure for RCTs assisted by a machine learning adjustment. Through a set of simulation experiments and a real-world example, we showed the proposed method remarkably boosts the statistical power and can reduce 1/3 sample size requirement as compared with other approaches. Such improvement is grounded in the capability of machine learning in capturing nonlinear relationships and higher order interactions, which cannot be fully exploited by linear models. We also demonstrated that the proposed method has proper type I error control. 
Meanwhile, as long as the outcome has a moderate level of dependency with the baseline covariates, the Wilcoxon-RF method never performs worse than the standard two-sample tests. Hence, the risk of using the proposed method is minimal, while the potential gain is substantial. In contrary, the methods using adjustment by linear regression can sometimes harm the efficiency, which can be attributed to its instability when $p$ is large relatively to $N$. Shrinkage methods such as Ridge \cite{hoerl1970ridge}, Lasso \cite{tibshirani1996regression} or Elastic net \cite{zou2005regularization} may be used, but they all require careful tuning of hyperparameters, and they do not address the difficulty brought by non-linearities and interactions.

The RF model is often considered an ``off-the-shelf" supervised learning method due to its simplicity of usage. Typically, with the recommended settings one can obtain a model with good performance \cite{hastie2009elements}. The usage of OOB observations further makes explicit training, validation set splitting unnecessary. Therefore, in an ideal scenario, the inference procedure can be completed in a single run of RF model training. This feature makes it particularly advantageous because the FDA has clear expectations that the analyses of primary and secondary endpoints in clinical trials need to be pre-specified. The post-hoc analyses are prone to data-driven manipulations which can lead to biased interpretation of the results. Therefore, any modeling method that is difficult to preplan and requires extensive tuning of hyperparameters can negatively affect the weight of evidence from an RCT for the regulatory decision-making. Another motivation for using RF is that it usually shows better or comparable prediction performance than neural networks in the range of sample sizes typically used in a randomized study. Deep learning models usually require larger sample size to gain advantage \cite{lecun2015deep}.  However, no model can be optimal under all scenarios. Although RF is used as a proof of concept in our study, other modeling methods can be superior to RF in many settings. Therefore, investigation of the proposed approach to incorporate other machine learning techniques is important.

Another practical challenge is how to translate the gain in efficiency into a reduction of the sample size. This requires the knowledge of the variation that can be explained by the machine learning model at the planning stage. In medical research, a phase II trial is conducted to preliminarily assess the efficacy before conducting a large scale phase III RCT. The data collected from the phase II trial provides important parameters that can guide the design of the phase III trial. One can simply use the two-sample $t$-test formula as an approximation for the exact test, and estimate the proportion of variance that will be reduced by including the covariates based on assumptions or data available, e.g. from a phase II trial. Approximately, if variance explained by the model is $\gamma$, then one would expect a reduction of $N\gamma$ in sample size requirement, where $N$ is the sample size needed using an unadjusted two-sample test. The parameters may also be used to run simulations to estimate the sample sizes required for the machine learning adjusted tests. Another option is an interim analysis on the efficiency gain and adjust the sample size accordingly. Such strategy may require further methodological investigation.

\bibliographystyle{unsrtnat}
\bibliography{reference}

\end{document}

\maketitle

\begin{figure}[h]
	\centering
	\includegraphics[width=7in]{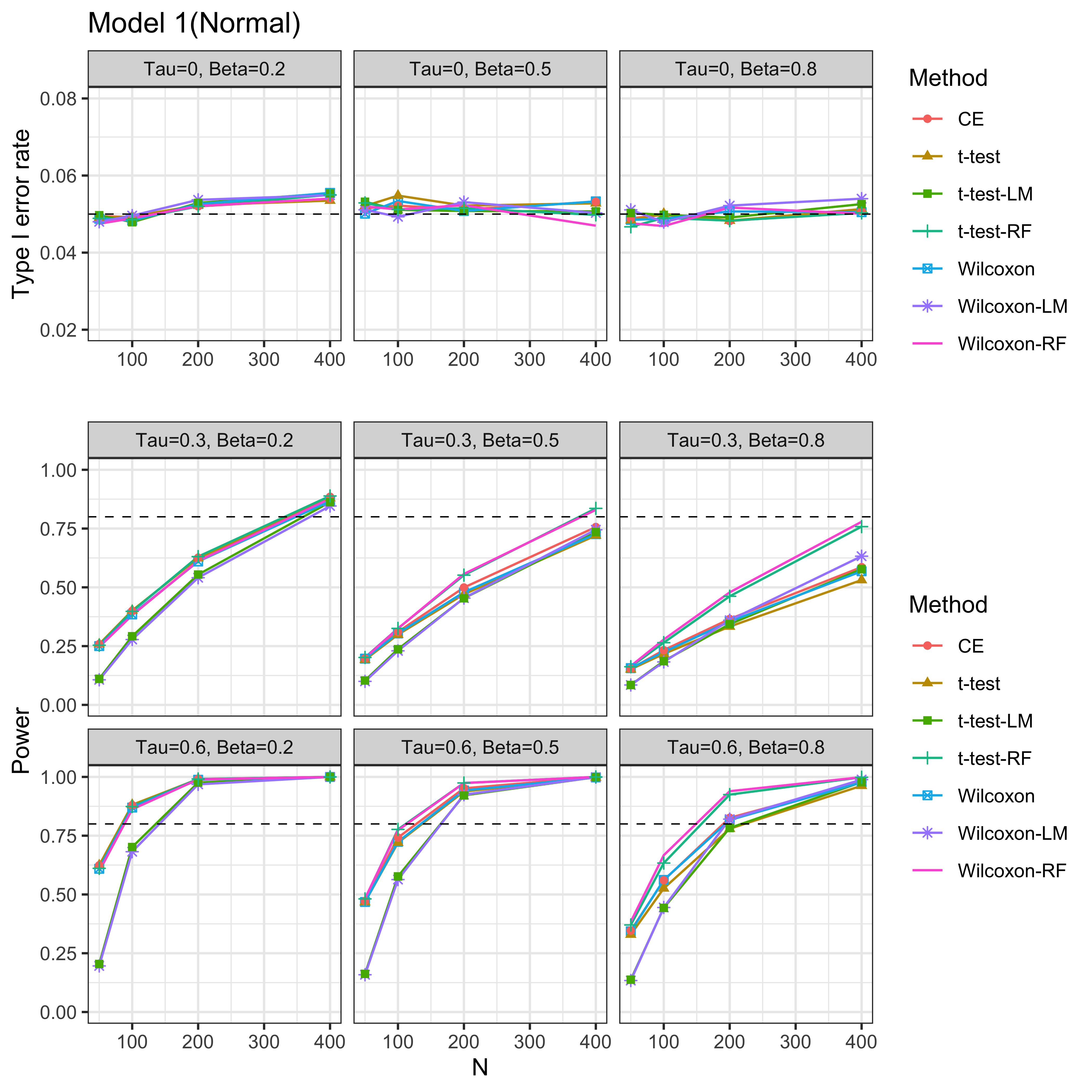}
	\caption{Type I error (Tau=0) and statistical power of the testing procedures under different settings of Model 1 and with error terms from standard log-normal  distributions.}
\end{figure}

\begin{figure}[h]
	\centering
	\includegraphics[width=7in]{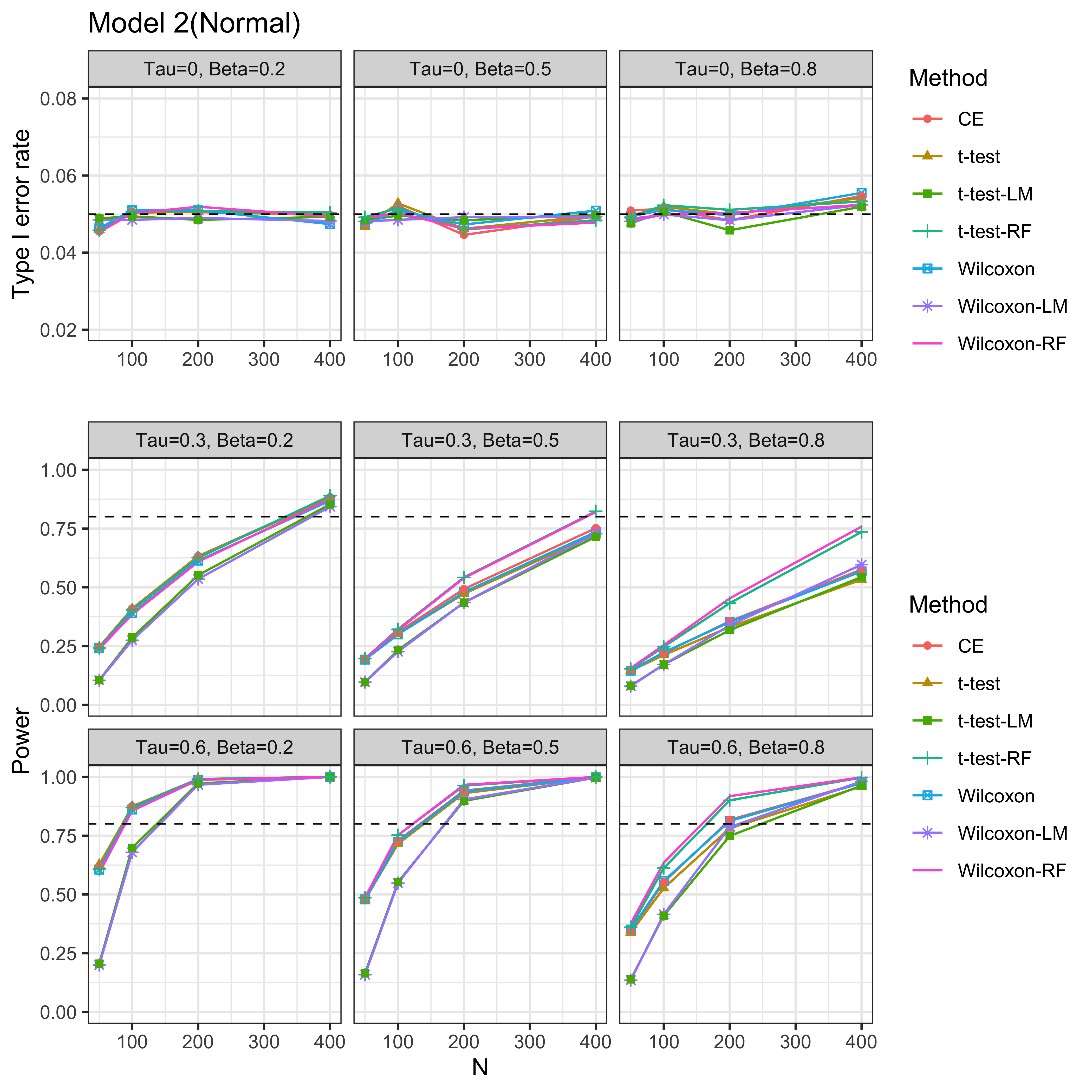}
	\caption{Type I error (Tau=0) and statistical power of the testing procedures under different settings of Model 2 and with error terms from standard log-normal  distributions.}
\end{figure}

\begin{figure}[h]
	\centering
	\includegraphics[width=7in]{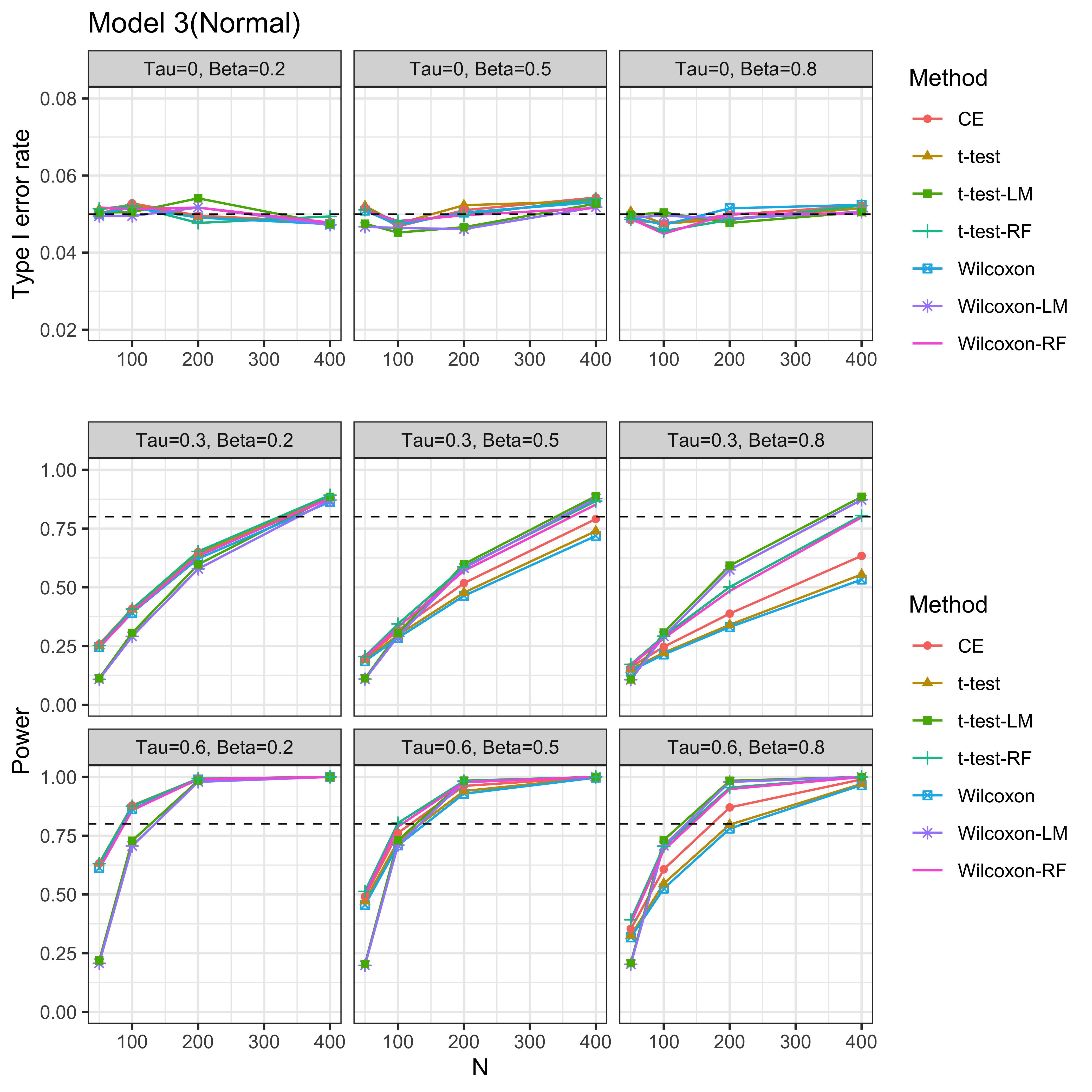}
	\caption{Type I error (Tau=0) and statistical power of the testing procedures under different settings of Model 3 and with error terms from standard log-normal  distributions.}
\end{figure}

\begin{figure}[h]
	\centering
	\includegraphics[width=7in]{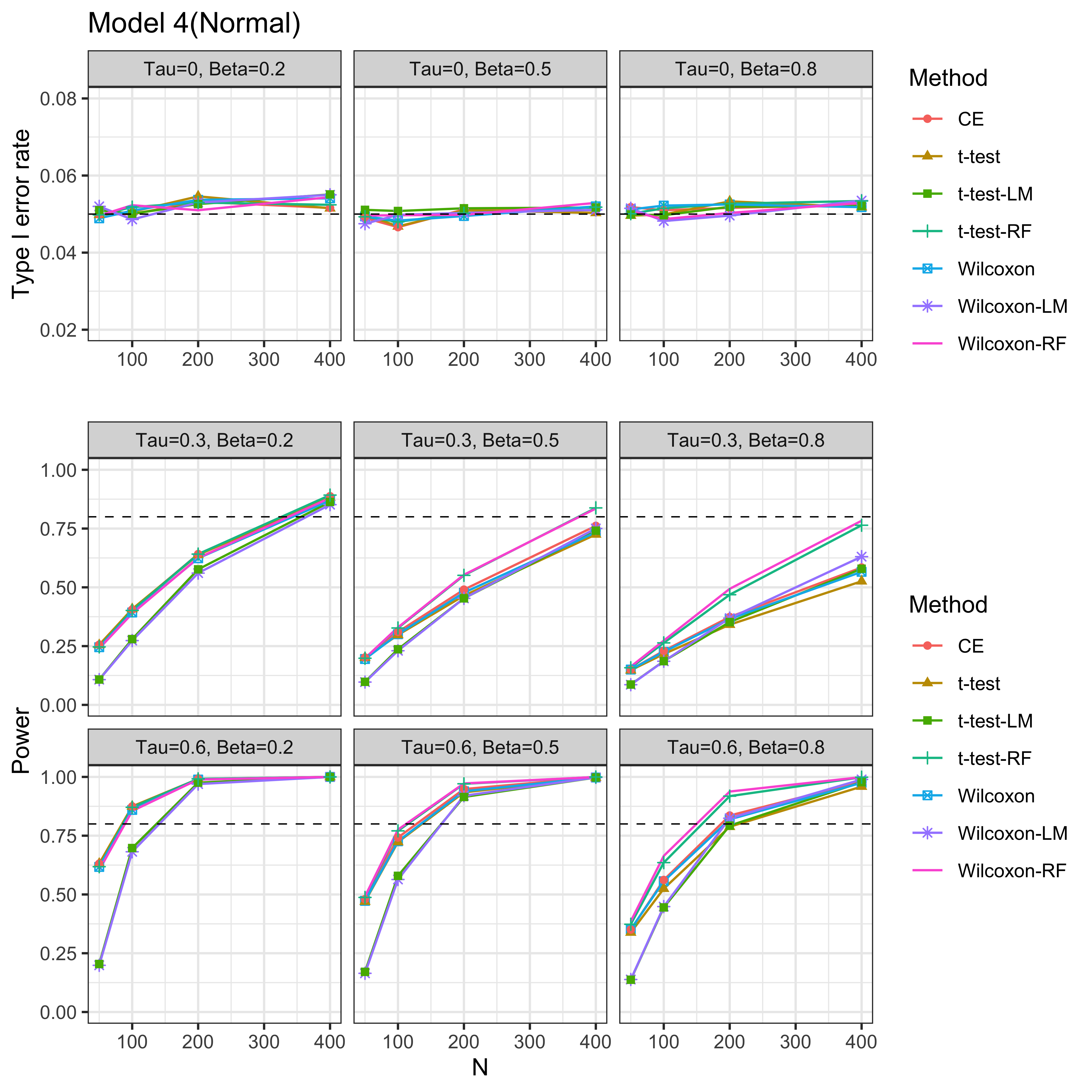}
	\caption{Type I error (Tau=0) and statistical power of the testing procedures under different settings of Model 4 and with error terms from standard log-normal distributions.}
\end{figure}

\begin{figure}[h]
	\centering
	\includegraphics[width=7in]{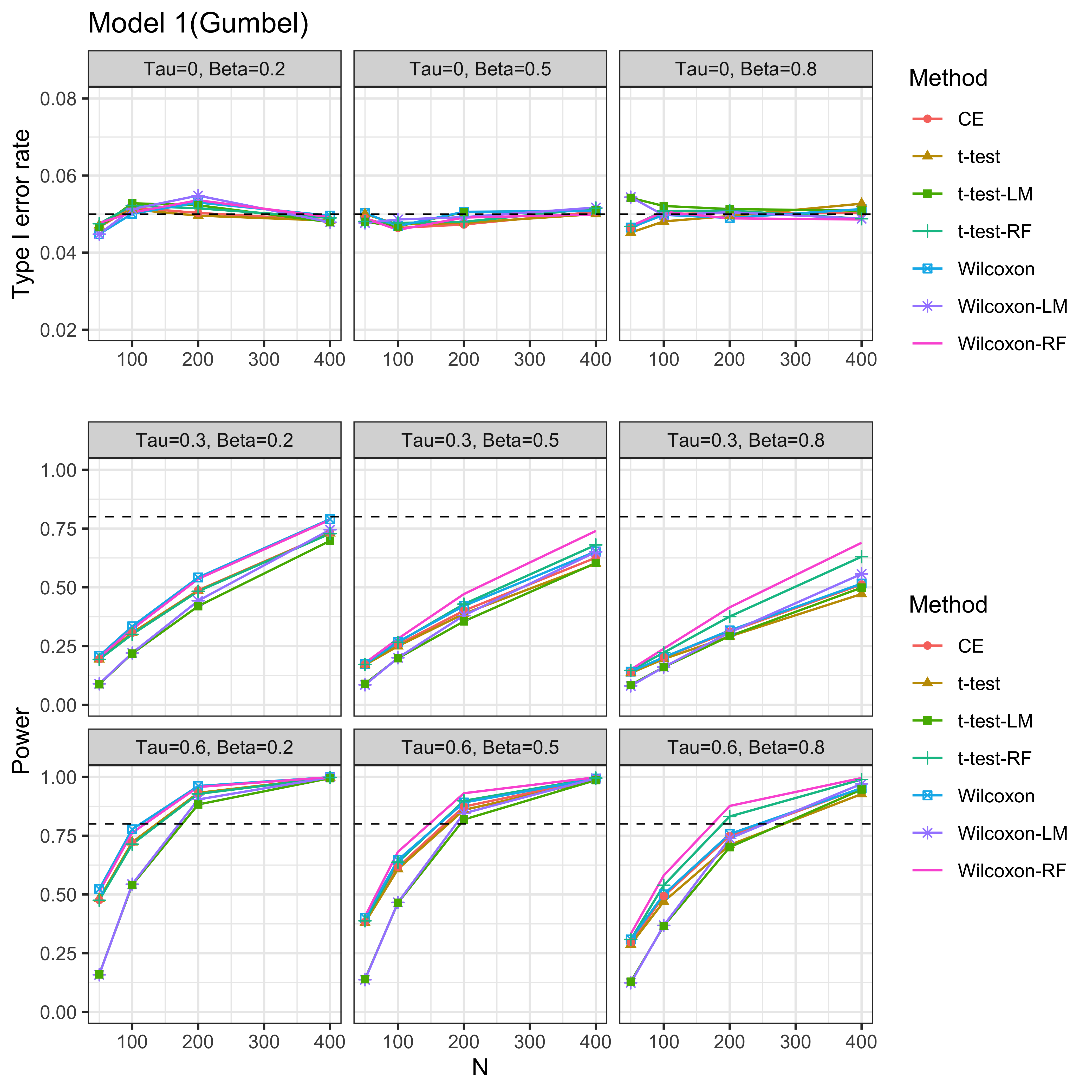}
	\caption{Type I error (Tau=0) and statistical power of the testing procedures under different settings of Model 2 and with error terms from standard Gumbel distributions.}
\end{figure}

\begin{figure}[h]
	\centering
	\includegraphics[width=7in]{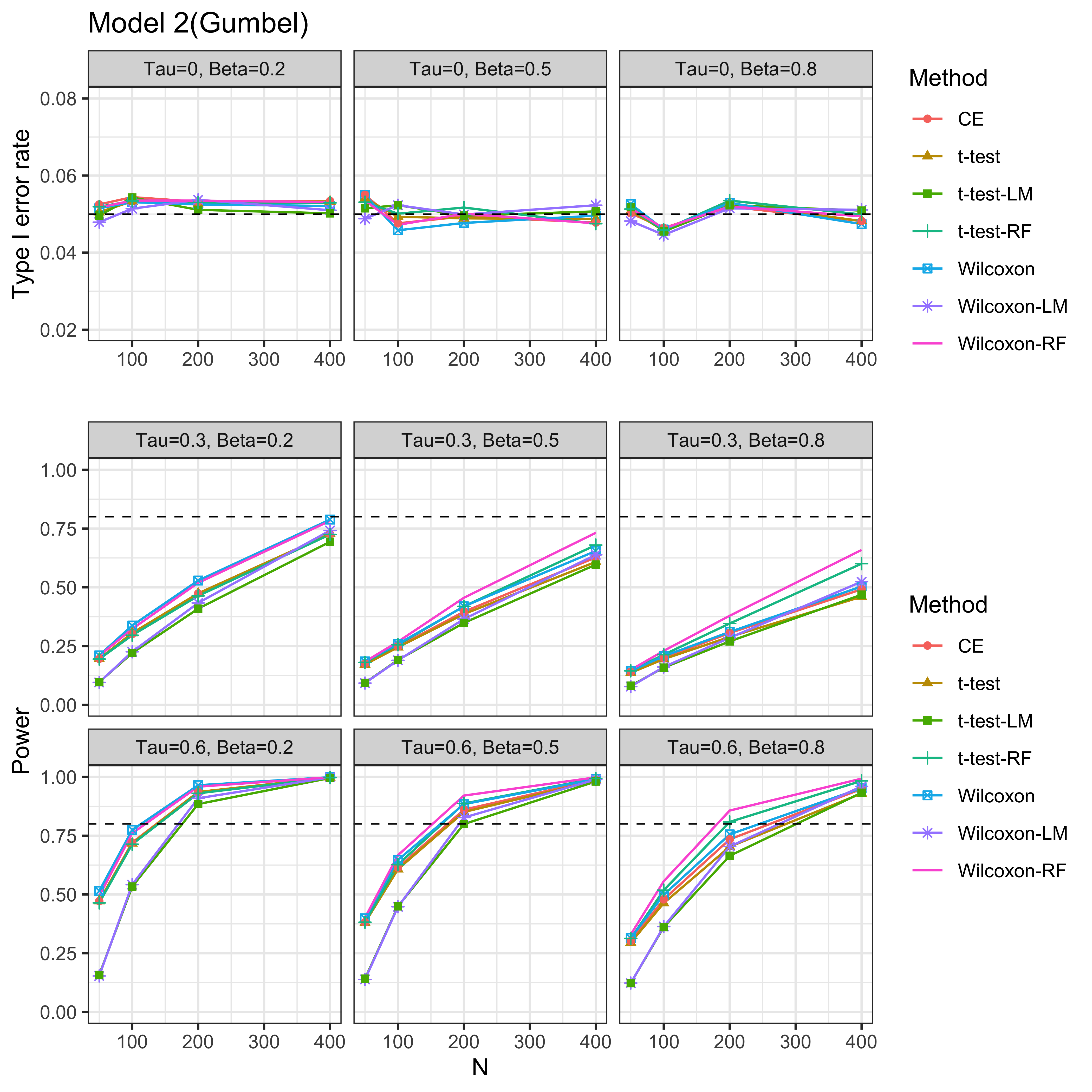}
	\caption{Type I error (Tau=0) and statistical power of the testing procedures under different settings of Model 2 and with error terms from standard Gumbel distributions.}
\end{figure}

\begin{figure}[h]
	\centering
	\includegraphics[width=7in]{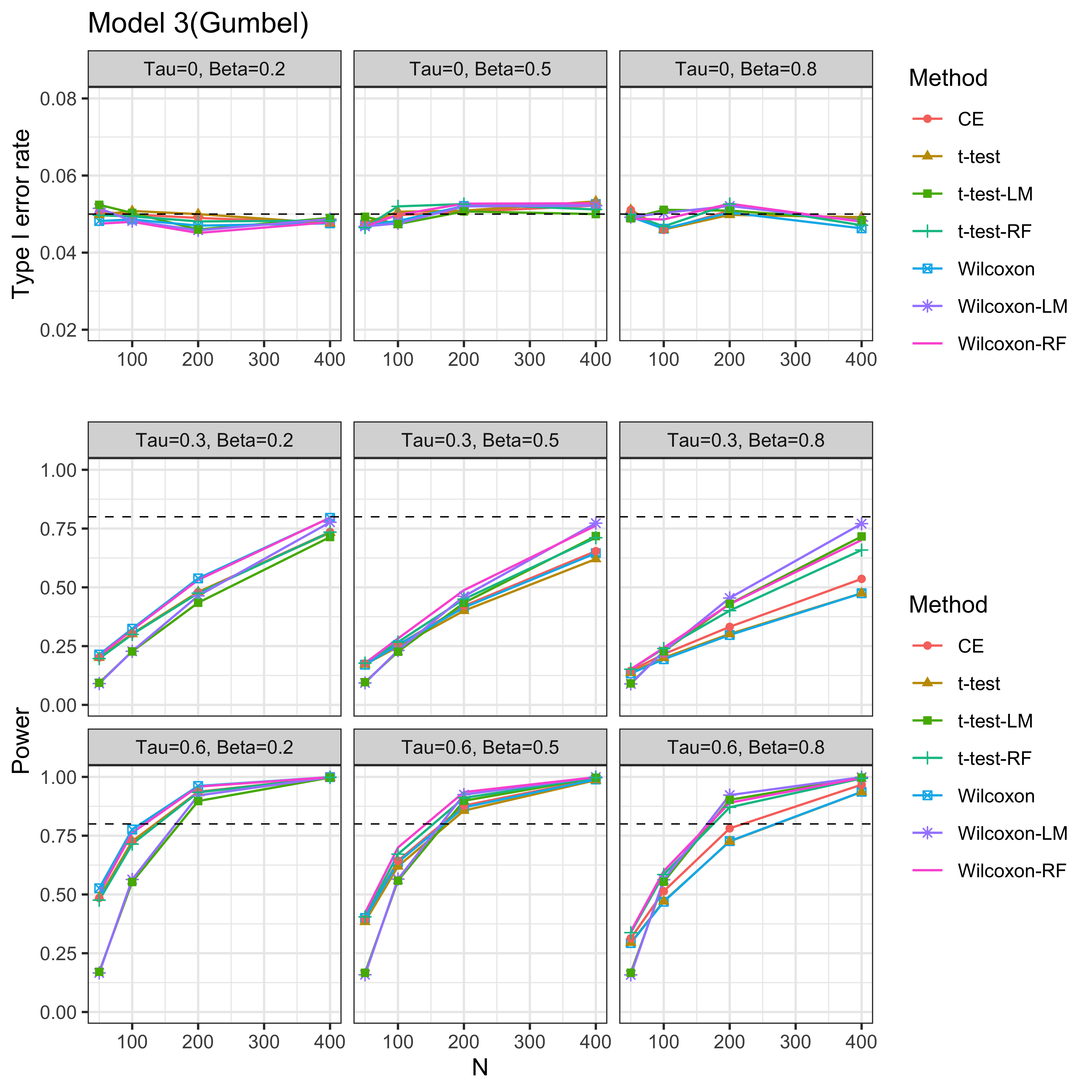}
	\caption{Type I error (Tau=0) and statistical power of the testing procedures under different settings of Model 3 and with error terms from standard Gumbel distributions.}
\end{figure}

\begin{figure}[h]
	\centering
	\includegraphics[width=7in]{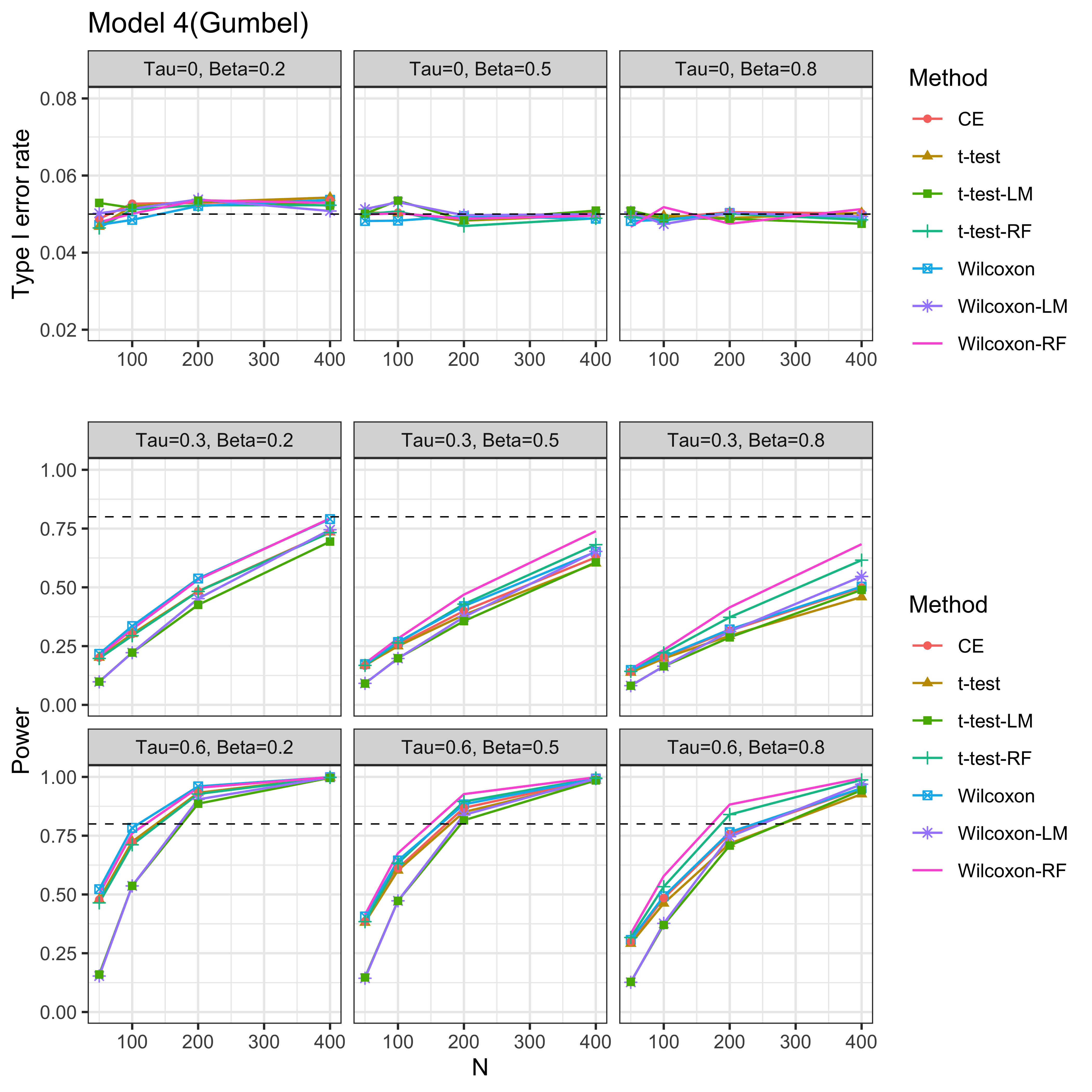}
	\caption{Type I error (Tau=0) and statistical power of the testing procedures under different settings of Model 4 and with error terms from standard Gumbel distributions.}
\end{figure}

\begin{figure}[h]
	\centering
	\includegraphics[width=7in]{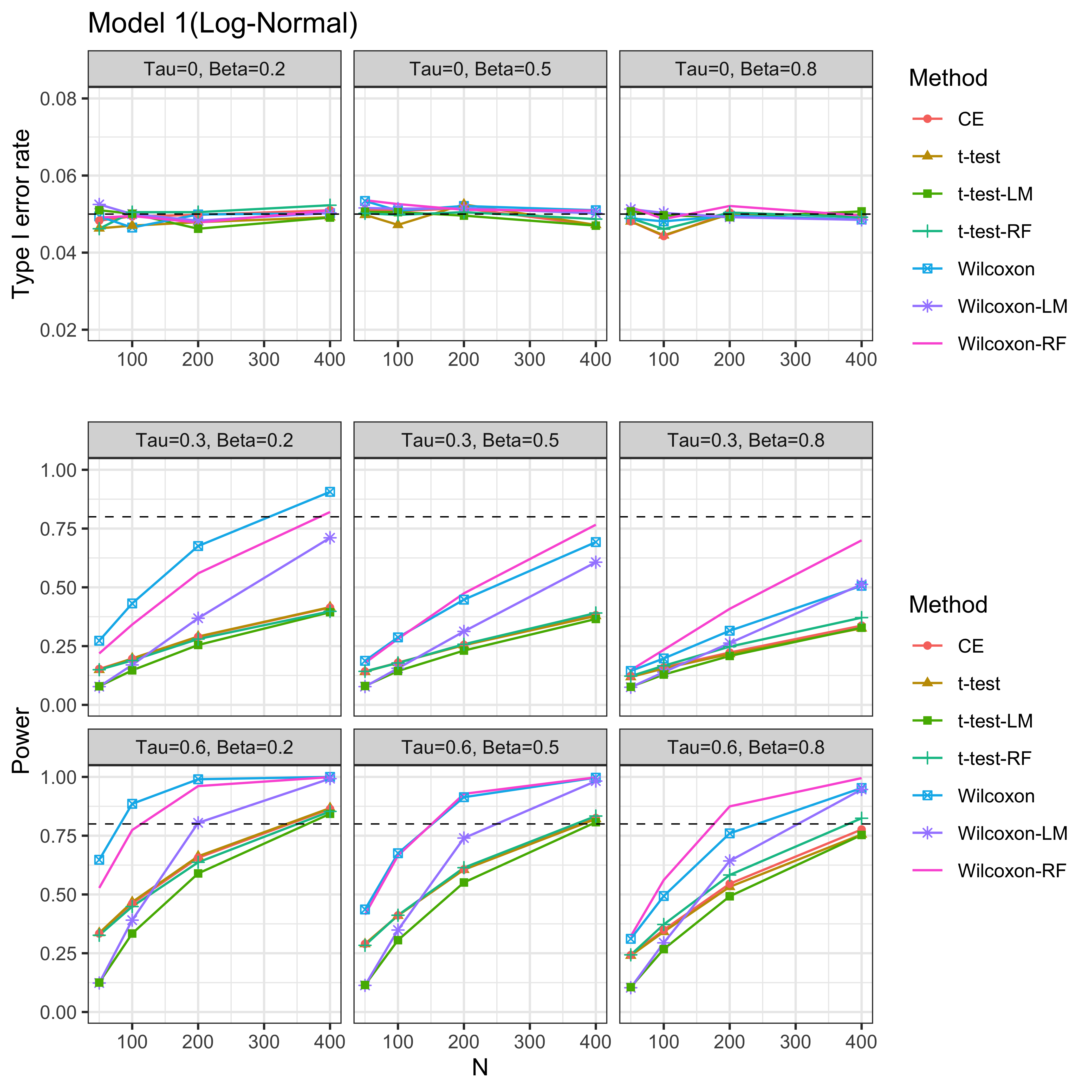}
	\caption{Type I error (Tau=0) and statistical power of the testing procedures under different settings of Model 1 and with error terms from standard log-normal  distributions.}
\end{figure}

\begin{figure}[h]
	\centering
	\includegraphics[width=7in]{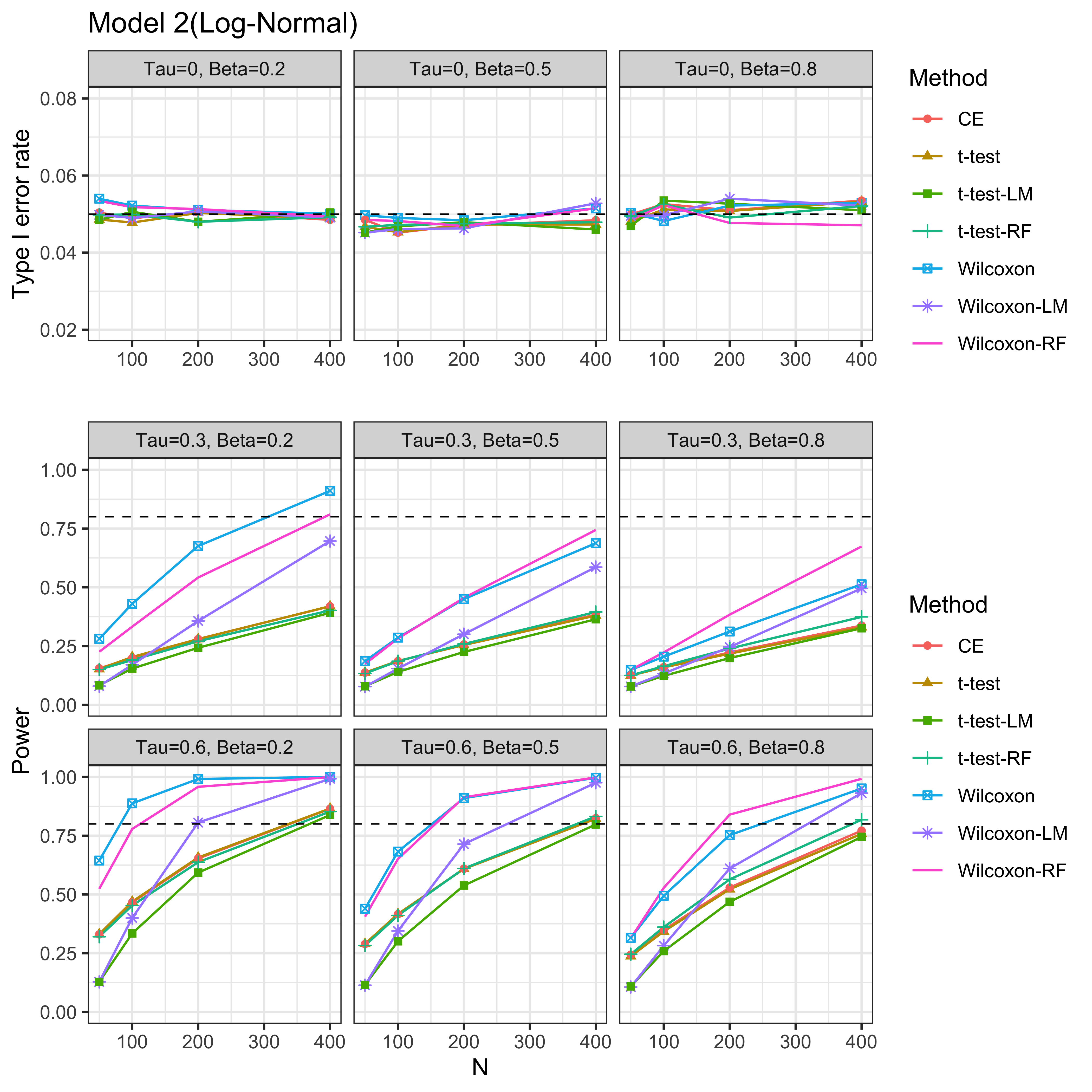}
	\caption{Type I error (Tau=0) and statistical power of the testing procedures under different settings of Model 2 and with error terms from standard log-normal  distributions.}
\end{figure}

\begin{figure}[h]
	\centering
	\includegraphics[width=7in]{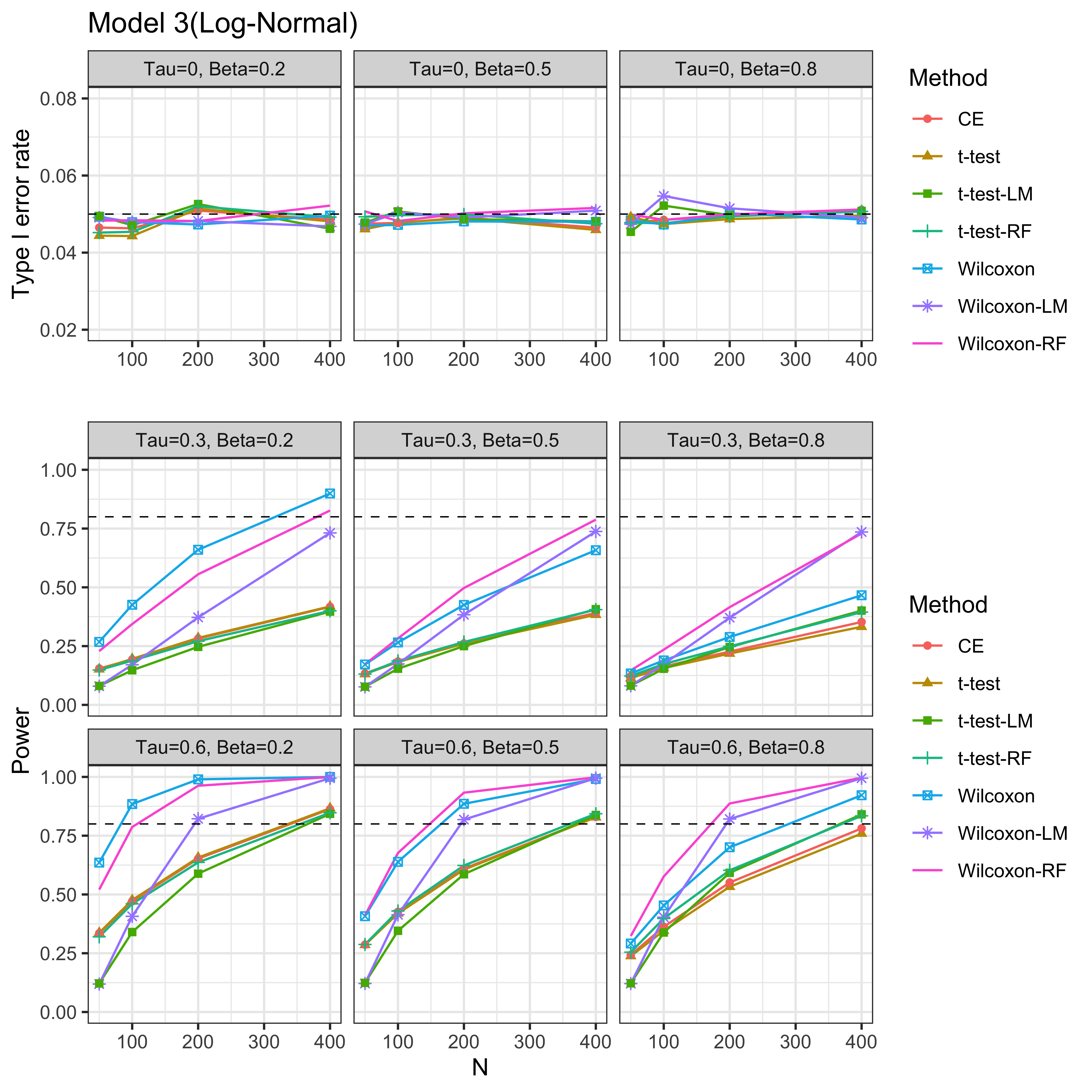}
	\caption{Type I error (Tau=0) and statistical power of the testing procedures under different settings of Model 3 and with error terms from standard log-normal  distributions.}
\end{figure}

\begin{figure}[h]
	\centering
	\includegraphics[width=7in]{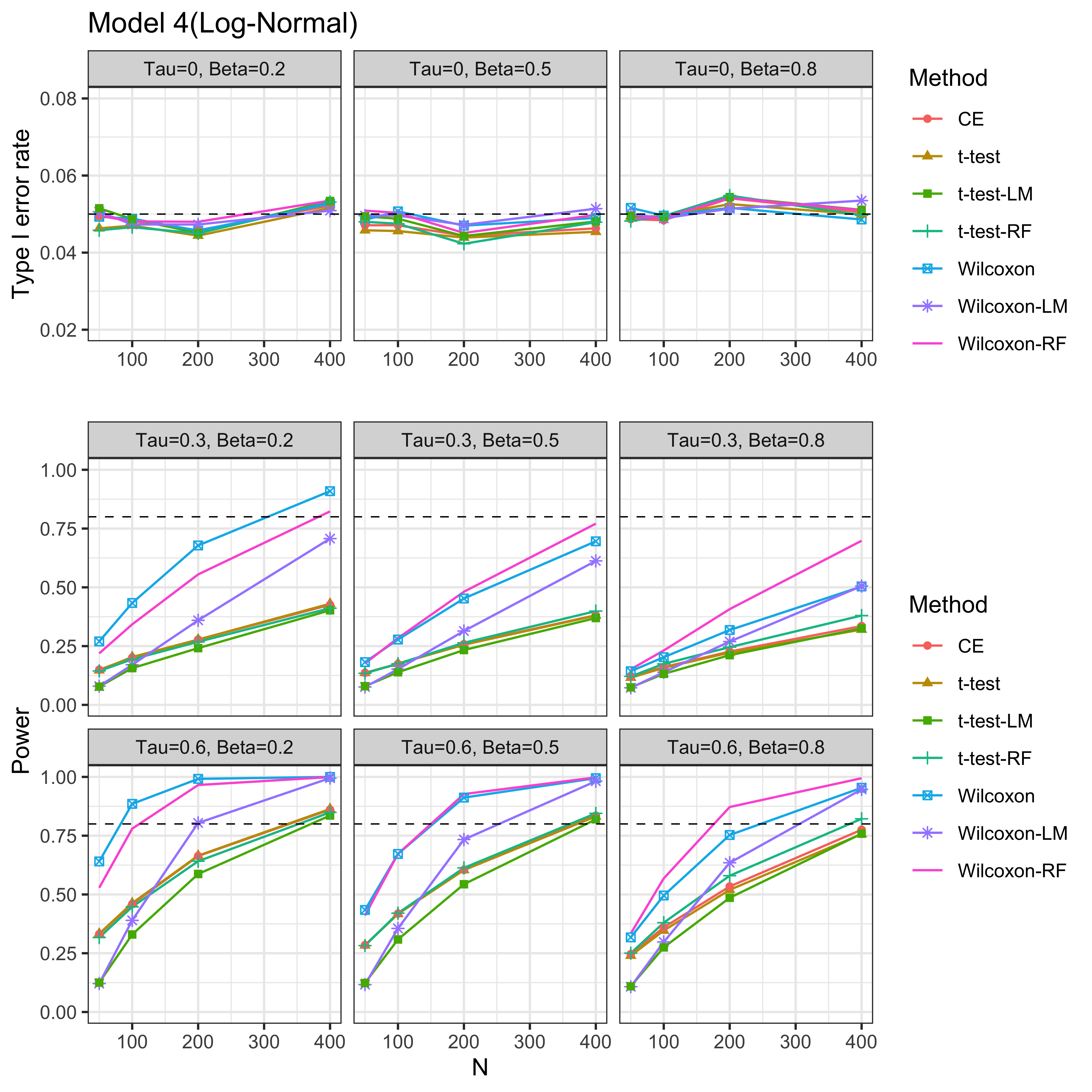}
	\caption{Type I error (Tau=0) and statistical power of the testing procedures under different settings of Model 4 and with error terms from standard log-normal distributions.}
\end{figure}